\documentclass[5p,twocolumn]{elsarticle}

\usepackage{graphicx}
\usepackage{dcolumn}
\usepackage{pifont}
\usepackage{bm}
\usepackage{multirow}
\usepackage{amsmath}
\usepackage{float}
\usepackage{subfig}
\usepackage{txfonts}
\usepackage[version=3]{mhchem}

\usepackage[usenames,dvipsnames]{xcolor}
\definecolor{blue}{RGB}{0,0,225}
\definecolor{cream}{RGB}{222,217,201}
\definecolor{red}{RGB}{225,0,0}

\journal{arXiv}

\begin{document}
\title{First-principles study on luminescence properties of Eu-doped defect pyrochlore oxide KNbWO$_6\cdot$H$_2$O:Eu$^{3+}$}

\author[kimuniv-m]{Song-Hyok Choe}
\author[kimuniv-m]{Chol-Jun Yu\corref{cor}}
\cortext[cor]{Corresponding author}
\ead{cj.yu@ryongnamsan.edu.kp}
\author[kimuniv-m]{Myong Choe}
\author[kimuniv-m]{Yun-Hyok Kye}
\author[kimuniv-c]{Yong-Nam Han}
\author[jilin]{Guangsheng Pang}

\address[kimuniv-m]{Chair of Computational Materials Design, Faculty of Materials Science, Kim Il Sung University, Ryongnam-Dong, Taesong District, Pyongyang, Democratic People's Republic of Korea}
\address[kimuniv-c]{Faculty of Chemistry, Kim Il Sung University, Ryongnam-Dong, Taesong District, Pyongyang, Democratic People's Republic of Korea}
\address[jilin]{State Key Laboratory of Inorganic Synthesis and Preparative Chemistry, College of Chemistry, Jilin University, Changchun 130012, P. R. China}

\begin{abstract}
Defect pyrochlore oxides have attracted a great interest as promising luminescent materials due to their flexible composition and high electron/hole mobility.
In this work, we investigate the structural and electronic properties of lanthanide-doped (Ln) defect pyrochlore oxides \ce{KNbWO6}:0.125Ln$^{3+}$ by using first-principles calculations.
We perform structural optimizations of various defect pyrochlore models and calculate their electronic structures, revealing that hydration has a significant influence on both local symmetry around Eu$^{3+}$ ion and band structures with an alteration of their luminescent behaviour.
In the hydrated compounds, the electric-dipole $^5$D$_0-^7$F$_2$ transition is found to be partially suppressed by the raised local symmetry, and the water molecules in the compounds can mediate the non-radiative energy transfer between the activator Eu$^{3+}$ ions and the host, resulting in the quenching effect.
It turns out that the oxygen vacancies are detrimental to luminescence as they reduce the Eu$^{3+}$ ion in its vicinity to Eu$^{2+}$ ion and also serve as traps for conduction electrons excited by incident light. 
Our calculations for \ce{KNbWO6}:0.125Ln$^{3+}$ (Ln = Ce, Pr, Nd, Pm, Sm) support that defect pyrochlore oxide \ce{KNbWO6} can also be used as luminescence host for Ln$^{3+}$ ion doping, giving a valuable insight into a variation trend in luminescent properties of these materials at atomic level.
\end{abstract}

\maketitle

\section{\label{intro}Introduction}
During the past half century, lots of luminescent materials have been discovered with numerous applications in many areas, including fluorescent lamp~\cite{jfang}, photovoltaics and photocatalyst~\cite{yang, xiaoyong, fengfeng}, bio-imaging~\cite{chen} and white light applications~\cite{mutlet}.
In order to improve efficiency, life-time and environmental friendliness, much effort has been devoted to develop advanced luminescent materials like lanthanide (Ln) ion doped oxides \ce{ZrO2}:Tm$^{3+}$, Tb$^{3+}$, Eu$^{3+}$~\cite{lovisa}, silicate nitrides \ce{LaSi3N5}:Ce$^{3+}$~\cite{gonze1} and phosphates \ce{LaPO4}:Eu$^{3+}$~\cite{jiyou}.
Among many other inorganic compounds used for luminescence hosts, the hollow crystal structures are desirable for accommodating the Ln ions with large ionic radii. 
In this sense, several pyrochlore compounds with large channel structure were explored as promising luminescent hosts~\cite{xinshuan, fujihara, gentleman}.

We synthesized the hydrated defect pyrochlore oxide \ce{KNbWO6.H2O} and identified the luminescence properties by experimental measurements~\cite{euHan, euHan2}.
Furthermore, Eu-doped hydrated defect pyrochlore \ce{KNbWO6.H2O}:$x$Eu$^{3+}$ was prepared by ion exchange under hydrothermal condition~\cite{euHan}.
In general, the Eu$^{3+}$ ions are known to exhibit characteristic red emission and to be excited by ultra violet (UV) light, and they are noticeable for symmetry-sensitive emission because of the non-degenerate ground state \ce{^7F0} and non-overlapping \ce{^{2s+1}L,J} multiplets.~\cite{biserka}
In \ce{KNbWO6.H2O}:$x$Eu$^{3+}$, the emission spectra with strong peaks at 580, 594 and 612 nm were observed, together with excitation spectra with peaks at 394 and 464 nm.
All these peaks in the emission/excitation spectra were identified as corresponding to intra-configurational $f-f$ transitions of $4f$ electrons of Eu$^{3+}$ ion.
The emission intensity was found to be the strongest at Eu-doping concentration of $x=0.131$, and to get much stronger after annealed at 450 $^\circ$C for 2 h. 
After annealing, the lifetime of emission also became significantly longer due to the effect of dehydration during annealing, implying that hydration might be a possible reason for luminescence quenching.
The dehydration of \ce{KNbWO6.H2O}:$x$Eu$^{3+}$ was also found to influence on the local symmetry of Eu$^{3+}$ ion, being evidenced from the fact that the intensity of emission peak at 612 nm, originated from the electric-dipole \ce{^5D0-^7F2} transition, was strengthened as much as 10 times after annealing, while the magnetic-dipole \ce{^5D0-^7F1} transition spectra at 594 nm got slightly stronger.
These results indicate that \ce{KNbWO6.H2O} can be used as efficient luminescent host.
In spite of such experimental findings, no theoretical or computational studies on this material have been found, and thus there is lack of the atomistic insight into its luminescent properties. 

The Ln-doped luminescence materials have been subject of many first-principles studies within density functional theory (DFT) framework.
In DFT calculations of the Ln-doped materials, where the luminescence properties are governed by transitions between the multi-electronic states of the activator Ln ions involving $4f$ electrons, the challenge is to consider the $4f$ electrons explicitly.
What is worse, as a single-particle ground-state theory, DFT cannot directly explain the transitions between the multi-electronic states.
In fact, excepting the constrained DFT (cDFT) method that can manipulate only the $4f-5d$ transition, mainly occurring in the Ce$^{3+}$ and Eu$^{2+}$ ions, with a computation of transition energy and Stokes shift~\cite{gonze3,gonze1,daicho}, it was found to be almost impossible for DFT calculation to reproduce experimentally observed spectra formed by intra-configurational $4f-4f$ transitions.
Nevertheless, DFT calculations have been generally accepted to provide sufficiently reliable information for crystalline structure and activator-ligand interaction in the Ln-doped materials, being important for estimating the trend in luminescence property.
Moreover, the DFT electronic structures can be used to determine the positions of activator levels relative to the host band edges, which are also important for finding optical host-activator combinations~\cite{maohuadu}.
The effect of defects in host compound on luminescence property can also be investigated with DFT calculations~\cite{cmMechanism, gonze6}.

In this study, we perform the DFT calculations on several models related to the Eu-doped hydrated pyrochlore oxide \ce{KNbWO6.H2O}:$x$Eu to investigate the effects of hydration and defects on its luminescence property at atomic level.
The optimized structures of various defect-containing models are determined by structural relaxations and subsequently their electronic structures are calculated, determining the position of $4f$ levels relative to the host band edges.
We discuss the potential use of these compounds as a luminescence host for doping other trivalent Ln ions.
In the remaining part of this paper, computational methods are given in Section~\ref{sec-method}, the results and discussion in Section~\ref{sec-result}, and the main conclusion in Section~\ref{sec-conc}.

\section{\label{sec-method}Methods}
\subsection{Structural models}
\begin{figure*}[!th]
\centering
\includegraphics[clip=true,scale=0.11]{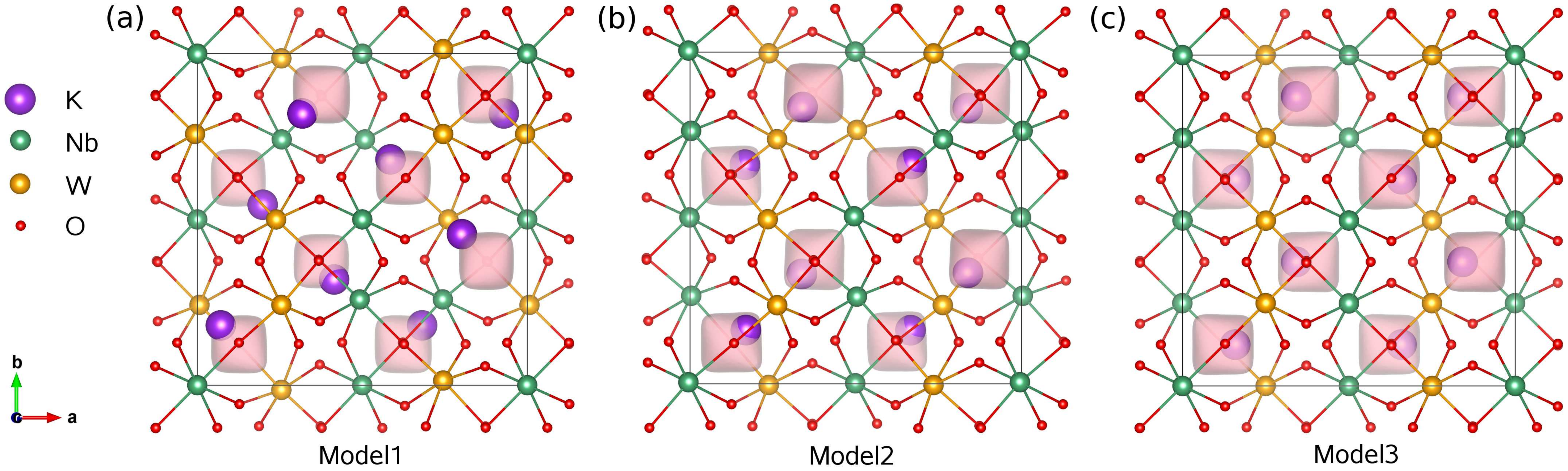}
\caption{Ball-and-stick view of unit cells optimized by using PBEsol + $U$ method for defect pyrochlore oxide \ce{KNbWO6} in three different models, distinguished by random distribution of Nb/W atoms; (a) disorder (model1), (b) semi-disorder (model2) and order (model3) models. Tetrahedron-like polyhedron with light pink colour indicates BVS isosurface at the value of BVS = 1 for K atom.}
\label{fig1}
\end{figure*}

The defect pyrochlore oxides with a chemical formula of \ce{AB2O6} were known to crystallize in cubic phase with a space group of \textit{Fd\={3}m}, where the crystalline lattices are composed of corner sharing \ce{BO6} octahedra network with 3D large channels and the A cations are located inside the channels.
The unit cell contains eight formula units (72 atoms).
In the case of \ce{KNbWO6}, the B sites are randomly occupied by Nb and W cations at the same probabilities, while the K cations randomly occupy one of every four equivalent crystallographic A sites.
We carried out bond valence sum (BVS) analysis to visualize the plausible locations of the K cations in relatively complex 3-dimensionally linked structural framework~\cite{Adam_BVS1, Adam_BVS3}, as shown in Fig. S1(a) where the isosurface at BVS = 1 for \ce{K+} ion is plotted with light pink colour.
From the BVS analysis, it is found that in the unit cell there are eight separate closed hollow spaces, resembling tetrahedron, inside which each \ce{K+} cation locates on any one of 4 vertex points, as depicted in Fig.~\ref{fig1}.
For the B sites, the configuration of Nb and W ions is truly random, and thus we suggested 3 different models distinguished by Nb/W sublattice ordering; model1 for disorder, model2 for semi-disorder and model3 for order of Nb and W sublattices.

When hydrating the pyrochlore oxide, the resultant \ce{KNbWO6$:$H2O} does not change the original crystalline lattice symmetry as the cubic phase with a space group of \textit{Fd\={3}m}, with a lattice constant of $a=10.5073$ \AA~and 8 formula units (96 atoms) in the unit cell~\cite{barnes}.
In this hydrated pyrochlore structure, the water molecules reside inside the hollow spaces and thus the K ions are pushed off the hollow (see Fig. S1(b)).
With a random placement of water molecule on any one of four equivalent hollow positions, we also suggested three different models for \ce{KNbWO6$:$H2O} crystal structure, distinguished by configuration of Nb/W random distribution on the B sites.
Meanwhile, for doping structures with the Eu ions, three \ce{K+} ions should be replaced by one \ce{Eu^{3+}} ion on the A sites to satisfy the electronic charge neutrality, leading to formation of one doped \ce{Eu^{3+}} ion and two potassium vacancies $V_{\text{K}}$.
Due to the sufficiently large size of lattice constants exceeding 10 \AA~and the number of atoms over 80, we can safely use the unit cell in the investigation of doped material rather than using the super cell.
Based on the experiment with a small amount of Eu atoms, only one Eu atom was supposed to be exchanged in the unit cell, leading to the compounds with chemical formula of \ce{KNbWO6}:0.125Eu$^{3+}$ for non-hydrated pyrochlore oxide and \ce{KNbWO6.H2O}:0.125Eu$^{3+}$ for hydrated one respectively.
We suggested three different configurations of Eu exchange for each non-hydrated model and two different configurations for the hydrated model.

\subsection{Computational details}
All the calculations in this study were carried out by means of pseudopotential plane-wave method within the DFT framework, as implemented in Quantum ESPRESSO (QE) package (version 6.2)~\cite{QE}.
As being available from the GBRV Library~\cite{gbrv}, the ultra-soft pseudopotentials (USPP) were used to describe the ion-electron interaction.
In the structural optimizations, the exchange-correlation interaction between valence electrons was considered by using the Perdew-Burke-Ernzerhof (PBE) functional~\cite{PBE} and its revised version for solids (PBEsol)~\cite{PBEsol} within generalized gradient approximation (GGA).
In addition, the Hubbard $U$~\cite{Cococcioni} term was taken into account for the $d$ states of Eu ($U=4$ eV), Nb and W ($U=3$ eV) atoms.
The wave functions and electronic densities were expanded by using the plane wave basis sets generated with the cut-off energies of 60 and 500 Ry, respectively.
Monkhorst-Pack~\cite{Monkpack} special \textit{k}-points for the Brillouin zone integration were set to be ($2\times2\times2$) for the structural optimization.
These computational parameters guarantee a total energy accuracy of 5 meV per formula unit.
The positions of all atoms and lattice constants were fully relaxed until the atomic forces converge to 0.02 eV/\AA.

In the electronic structure calculations including the energy bands and density of states (DOS), we only used the PBEsol functional, since it could give the better result than PBE when compared with the experimental lattice constants.
Spin-polarization was considered and the denser \textit{k}-points of ($4\times4\times4$) were used.
For the doped models, the $4f$ electrons of Eu were explicitly treated as the valence electrons by using USPP from PS Library (version 1.04)~\cite{pslibrary}, where the valence electron configuration is $5s^2 6s^2 5p^6 4d^{10} 5d^{0.5} 4f^{6.5}$.
In such $4f$ states-explicit calculations, the larger cut-off energies of 90 and 800 Ry were used for the sake of convergence.
The other doped models with Ln ions (Ce, Pr, Nd, Pm, Sm) were also investigated by using the same computational parameters to the case of Eu doping.

\section{\label{sec-result}Results and discussion}
\subsection{Structural property}
First, we carried out structural optimization of the unit cells for 3 different \ce{KNbWO6} models by using the PBE and PBEsol XC functionals.
By comparing with the experimental lattice constant, the result can be used as a check on a validity of the computational parameters and XC functionals.
Table~\ref{tab1} present the obtained lattice constants and volume of the unit cells for the three different models in comparison with the experimental one.
We note that when allowing the full relaxation of lattice parameters and atomic positions, the optimized crystal structures deviate from the cubic system as obtaining the different lattice constants of $a\neq b$ or $a\neq c$ or $b\neq c$, which can be thought to be caused by somewhat artificial (random) distribution of the atoms.
In this situation, the unit cell volume was also provided to make it easy to do comparison with experiment.
It turns out that when compared with the experiment, the PBE functional yielded an overestimation of lattice constants with relative errors of 3.1\%, 2.4\% and 2.2\% for model1, model2 and model3, whereas the PBEsol functional provided a good estimation of lattice constants with much lower relative errors of 0.7\%, 0.1\% and 0.5\% respectively.
This indicates that the PBEsol functional is more adequate for the defect pyrochlore oxides than PBE and the computational parameters used in this work are sufficiently reliable to give their accurate material properties.
Therefore, only the PBEsol functional will be used in the following calculations.
\begin{table}[!th]
\small
\caption{Lattice constant ($a$, $b$, $c$), unit cell volume and total energy differences in three different \ce{KNbWO6} models shown in Fig.~\ref{fig1}, calculated with PBE and PBEsol XC functionals.}
\label{tab1}
\begin{tabular}{lccccc}
\hline
\multirow{2}{*}{Model} & \multicolumn{3}{c}{Lattice constants (\AA)} & Volume & $\Delta E$ \\ 
\cline{2-4}
& $a$ & $b$ & $c$ & (\AA$^3$) & (meV/atom) \\
\hline
\multicolumn{4}{c}{PBE} & & \\
Model1 & 10.44 & 10.51 & 10.44 & 1147.50 & 0.00 \\
Model2 & 10.43 & 10.43 & 10.47 & 1139.72 & 0.09 \\
Model3 & 10.51 & 10.40 & 10.40 & 1137.62 & 5.21 \\
\hline
\multicolumn{4}{c}{PBEsol} & & \\
Model1 & 10.37 & 10.41 & 10.37 & 1120.46 & 0.00 \\
Model2 & 10.35 & 10.35 & 10.38 & 1113.92 & 2.22 \\
Model3 & 10.38 & 10.32 & 10.32 & 1107.61 & 7.03 \\		
\hline
Exp.~\cite{origStr} & \multicolumn{3}{c}{$a=b=c=10.36$} & 1112.84 & $-$ \\
\hline
\end{tabular}
\end{table}

In order to pick out the most stable model among the three different models, we present their total energy differences with respect to the lowest one in Table~\ref{tab1}.
The model1, disorder model for Nb/W distribution on the B sites, was found to have the lowest total energy, while the model3, order model for Nb/W sublattices, to have the highest total energy with the total energy differences of 5.21 and 7.03 meV/atom in the PBE and PBEsol calculations respectively.
It is worth noting that such tendency in total energy is consistent well with the disorder degree of Nb/W distribution on the B sites and can be associated with a strength of interaction between the \ce{K+} cations and the oxygen anions.
In fact, as shown in Fig.~\ref{fig1} for the optimized atomistic structures of the unit cells for the three different models, the \ce{K+} cations reside at the edge of BVS tetrahedron in the model1, formed by isosurface value of BVS = 1 that displays the hollow spaces for the \ce{K+} cations, while in the model3 they locate a little inside the BVS tetrahedron.
Therefore, the distance between the \ce{K+} cations and the neighboring oxygen anions becomes longer going from the model1 to the model2 and to model3, indicating a weakening of the K$-$O interaction and thus the stability of compound.

\begin{figure*}[!th]
\centering
\includegraphics[clip=true,scale=0.11]{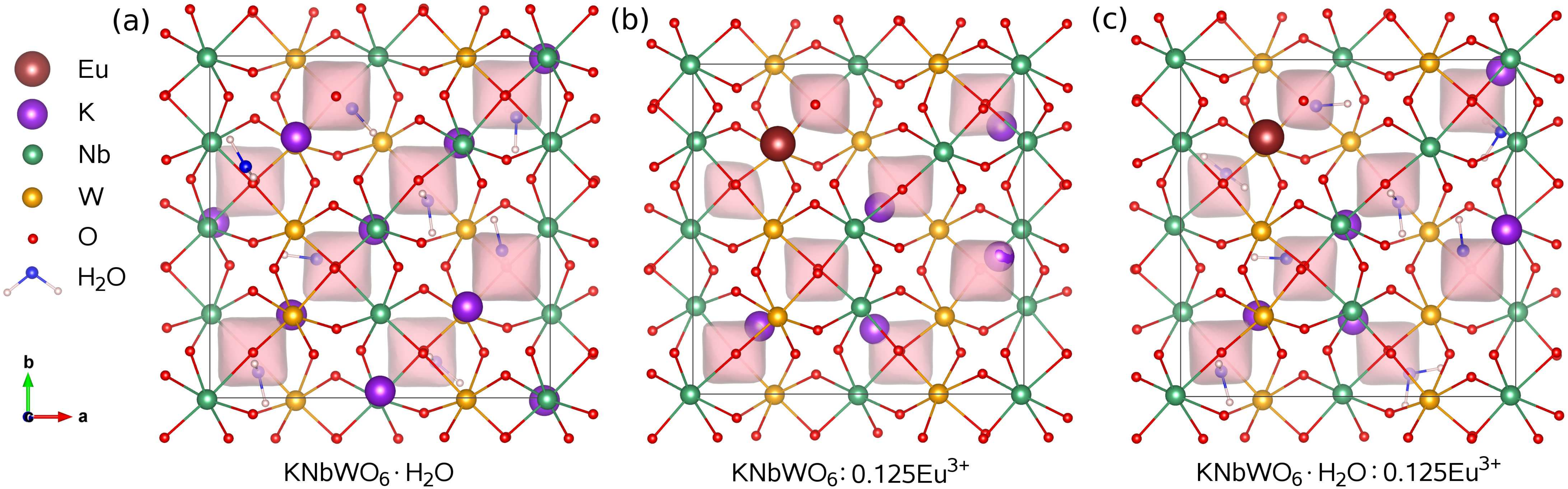}
\caption{Ball-and-stick view of unit cells with the lowest total energy for (a) hydrated defect pyrochlore oxide \ce{KNbWO6.H2O}, (b) Eu-doped pyrochlore oxide \ce{KNbWO6}:0.125Eu$^{3+}$ and (c) hydrated Eu-doped pyrochlore oxide \ce{KNbWO6.H2O}:0.125Eu$^{3+}$, optimized by using PBEsol + $U$ method. Tetrahedron-like polyhedron with light pink colour indicates BVS isosurface at the value of BVS = 1 for K atom.}
\label{fig2}
\end{figure*}
Using the PBEsol functional, we then carried out structural optimization of the unit cells for the hydrated pyrochlore \ce{KNbWO6.H2O} with 3 different models (model1, 2, 3) and for the Eu-doped pyrochlore \ce{KNbWO6}:0.125Eu$^{3+}$ with 9 different models (model1-1, 1-2, 1-3, 2-1, $\ldots$ , 3-3).
The obtained lattice constants and total energy differences of all the unit cells are presented in Table S1.
Unlike the original pyrochlore oxide, the lowest energy models were found to be based on model2 (semi-disorder model) for the hydrated and doped pyrochlore oxide.
For the Eu-doped hydrated pyrochlore \ce{KNbWO6.H2O}:0.125Eu$^{3+}$, therefore, we constructed the two different unit cells of Eu exchange based on the optimized model2 of \ce{KNbWO6.H2O} (Table S1).
In Table~\ref{tab2}, we summarize the optimized lattice constants and unit cell volumes of the energetically most favourable models for the original, hydrated, doped and hydrated-doped pyrochlore compounds, obtained by using the PBEsol functional, in comparison with the available experimental data for unit cell volume.
For brevity, the original \ce{KNbWO6} is termed as Orig, the hydrated \ce{KNbWO6.H2O} as Hyd, the Eu-doped\ce{KNbWO6}:0.125Eu$^{3+}$ as Dop and the hydrated Eu-doped \ce{KNbWO6.H2O}:0.125Eu$^{3+}$ as HydDop in the following.
The optimized unit cell volumes can be said to be in good agreement with the experimental values as their relative errors of 0.68\% in Orig, 0.92\% in Hyd and 0.50\% in HydDop models, again indicating a reliability of our models and computational method.
It was found that the volume expands with hydration while shrinks with \ce{Eu} doping.

\begin{table}[!b]
\small
\caption{Lattice constants ($a$, $b$, $c$) and unit cell volume of the most stable models for \ce{KNbWO6} (denoted as Orig), \ce{KNbWO6.H2O} (Hyd), \ce{KNbWO6}:0.125Eu$^{3+}$ (Dop) and \ce{KNbWO6.H2O}:0.125Eu$^{3+}$ (HydDop), calculated with PBEsol functional and $U$ method.}
\label{tab2}
\begin{tabular}{lcccccc}
\hline
\multirow{2}{*}{Model} & \multicolumn{3}{c}{Lattice constants (\AA)} & & \multicolumn{2}{c}{Volume (\AA$^3$)} \\
\cline{2-4} \cline{6-7}
& $a$ & $b$ & $c$ && Calc. & Exp. \\
\hline
Orig  & 10.37 & 10.41 & 10.37 && 1120.46 & 1112.84~\cite{origStr} \\
Hyd   & 10.50 & 10.52 & 10.39 && 1149.32 & 1160.04~\cite{barnes} \\
Dop   & 10.38 & 10.33 & 10.36 && 1111.31 & $-$ \\
HydDop& 10.45 & 10.50 & 10.39 && 1138.72 & 1144.44~\cite{euHan} \\
\hline
\end{tabular}
\end{table}

Figure~\ref{fig2} shows the crystalline structures of Hyd, Dop and HydDop models in ball-and-stick view.
When hydrating the original pyrochlore oxide \ce{KNbWO6}, water molecules were found to penetrate into the hollow spaces where the \ce{K+} cations already reside, resulting in the volume expansion, as shown in Fig.~\ref{fig2}(a).
One can see that the water molecules are located on the inner side of the hollow spaces and thus the \ce{K+} cations move to the outer side when compared with the non-hydrated pyrochlore, getting closer to the neighboring oxygen atoms.
On the other hand, the \ce{K+} cations in the Dop model maintain their locations as in the Orig model, while the inserted \ce{Eu^{3+}} ion is clearly away from the BVS tetrahedron, as shown in Fig.~\ref{fig2}(b) and (c).
In fact, due to its higher valence, \ce{Eu^{3+}} cation tends to interact more strongly with the neighboring \ce{O^{2-}} anions than \ce{K+} cation, so that it resides in the middle position between the hollow spaces.
It should be noted that such strengthening of the Eu$-$O binding leads to the contraction of volume when doping Eu into the defect pyrochlore oxide.
For the case of HydDop model shown in Fig.~\ref{fig2}(c), the location of \ce{Eu^{3+}} cation was found to be similar to the case of Dop model and \ce{K+} cations to be located on similar position to the case of Hyd model.

We further consider local symmetry of the \ce{Eu^{3+}} ion in the hosts, which plays a critical role in luminescence as it determines the emission wavelength and intensity due to the intra-configurational $4f-4f$ transition~\cite{mahesh1, mahesh2}.
It is well known that for the case of \ce{Eu^{3+}} ion being at an inversion symmetry site, the electric dipole \ce{^5D0}--\ce{^7F2} transition is parity forbidden, while the magnetic dipole \ce{^5D0}--\ce{^7F1} transition is parity allowed with dominant emission wavelength around 590 nm.
Meanwhile, at a non-inversion symmetry site, the \ce{Eu^{3+}} ion is known to exhibit electric dipole \ce{^5D0}--\ce{^7F2} transition with emission wavelength of $610-620$ nm, of which intensity is hypersensitive to the site symmetry of \ce{Eu^{3+}}~\cite{fujihara}.
In several Eu-doped pyrochlore oxides with luminescence, both the electric and magnetic dipole transitions appear simultaneously, one of which is much stronger in most cases~\cite{fujihara,gentleman,mahesh1,mahesh2}.
In the case of defect pyrochlore oxide \ce{KNbWO6}, the location of doped \ce{Eu^{3+}} could be a highly symmetric site with inversion symmetry for ordering distribution of Nb/W on the B sites (see Fig. S2).
For truly random distribution of Nb/W atoms, however, the local symmetry of \ce{Eu^{3+}} is lowered, so that the electric dipole \ce{^5D0}--\ce{^7F2} transition as well as the magnetic dipole \ce{^5D0}--\ce{^7F1} transition can be observed~\cite{euHan}.
In addition, the random location of surrounding \ce{K+} ions inside the hollow spaces, due to their fractional occupancy of $1/4$, can also lower the site symmetry of \ce{Eu^{3+}} ion.
In this context, one can see that due to existence of water molecules in the hollow spaces, the distribution of \ce{K+} ions in the HydDop model is more symmetric than in the Dop model.
Therefore, the hypersensitive electric dipole \ce{^5D0}--\ce{^7F2} transition is more likely in the dehydrated Dop model, in qualitative agreement with the experimental observation~\cite{euHan}.

\subsection{Electronic structure}
\begin{figure}[!b]
\centering
~\includegraphics[clip=true,scale=0.5]{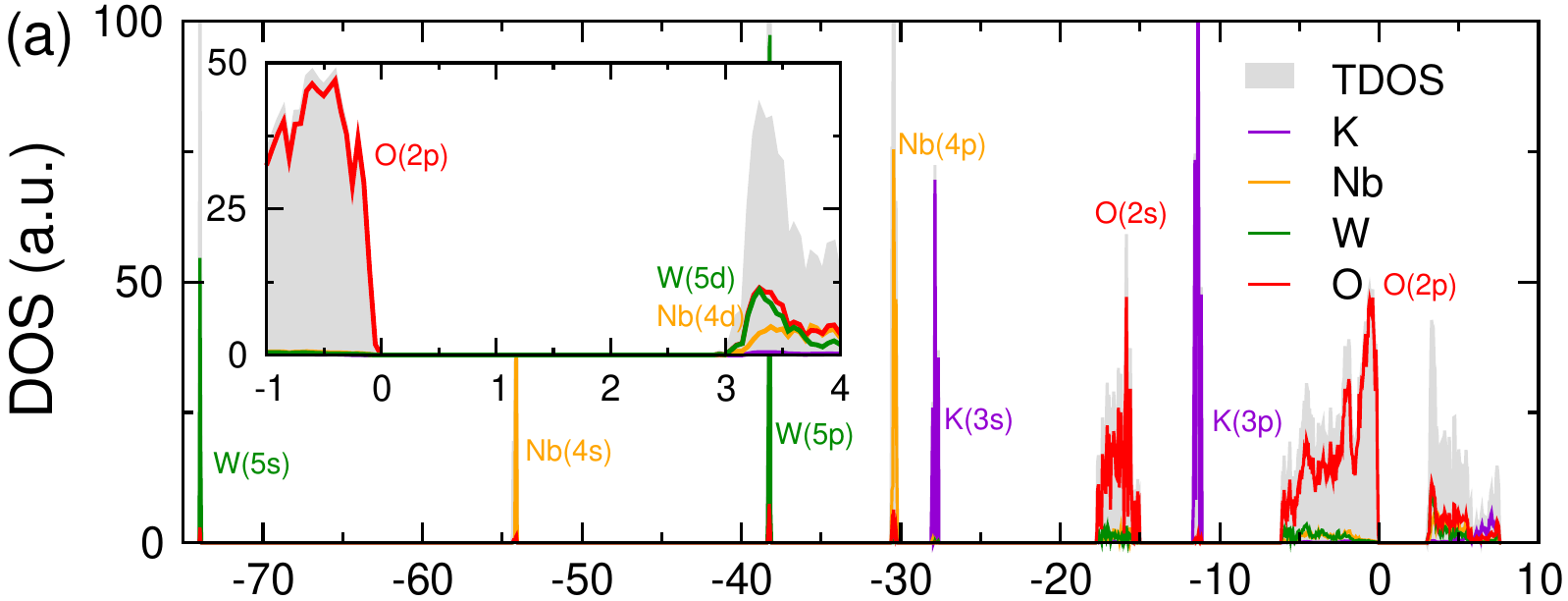}
\includegraphics[clip=true,scale=0.5]{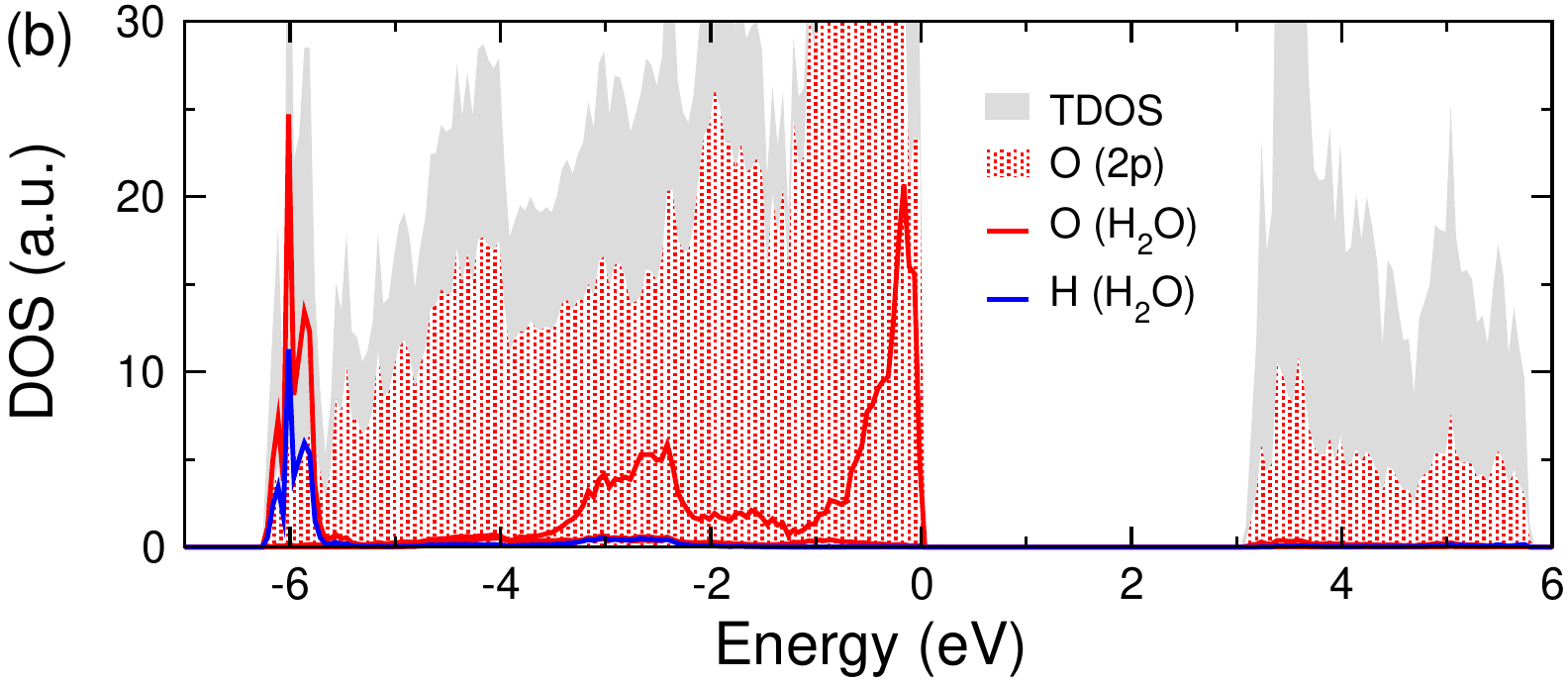}
\caption{Density of states (DOS) for (a) defect pyrochlore oxide \ce{KNbWO6} (Orig model) in the full energy range from $-$75 eV to 10 eV, where the inset shows DOS around the Fermi level in detail, and (b) hydrated defect pyrochlore oxide \ce{KNbWO6.H2O} (Hyd model) in the energy range from $-$7 eV to 6 eV, calculated by using PBEsol + $U$ method.}
\label{fig3}
\end{figure}
Using the optimized models, i.e. Orig, Hyd, Dop and HydDop models, with the lowest total energies, we calculated their electronic structures by means of PBEsol + $U$ method to investigate the luminescent properties.
First, those of the undoped Orig and Hyd models are discussed.
In these models, the band structures for spin-up and spin-down states were found to be identical each other, and thus those for spin-up states were only shown in Fig. S3.
The band gaps were calculated to be 3.00 eV for the defect pyrochlore oxide \ce{KNbWO6} (Orig) and 3.03 eV for the hydrated \ce{KNbWO6.H2O} (Hyd), which can be said to be in reasonable agreement with the experimental value of 3.50 eV~\cite{Zeng18njc} compared with other GGA calculations for semiconductors, indicating the suitability of our selected $U$ values.
The almost identical band gap of the hydrated pyrochlore oxide with the original dehydrated one indicates that the water molecules inside the crystal hardly affect the electronic structure of host material.
Figure~\ref{fig3} displays the density of states (DOS) for the spin-up states, projected on atomic orbitals.
In both the cases of Orig and Hyd models, the valence bands (VBs) were found to be dominantly originated from the O $2p$ states, while the conduction bands (CBs) to be composed of the $4d/5d$ states of Nb/W atoms and the O $2p$ states.
For the case of Hyd model, the electronic states of O and H atoms of water molecule appeared in VBs, being localized compared with the O $2p$ states of host.
None of states of water molecule are observed in CBs, again indicating the relatively weak interaction between water molecules and host material.

\begin{figure}[!t]
\centering
\includegraphics[clip=true,scale=0.18]{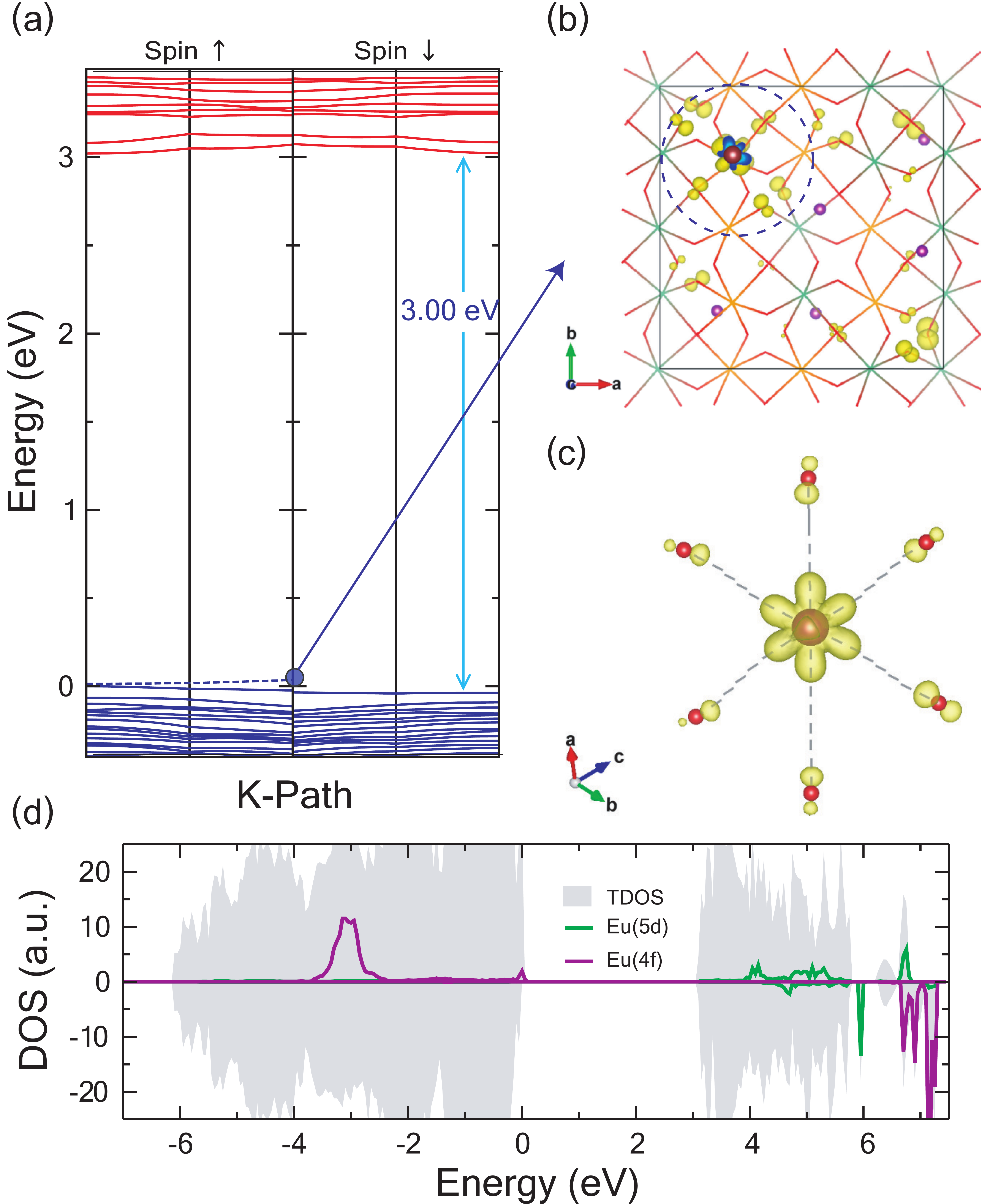}
\caption{Electronic structure of Eu-doped defect pyrochlore oxide \ce{KNbWO6}:0.125\ce{Eu^{3+}} (Dop model), calculated with PBEsol + $U$ method. (a) Energy band structure for spin-up and spin-down sates with dotted line for unoccupied band and unchanged band gap of 3.00 eV, (b) isosurface view of wave function square at $\Gamma$ point, corresponding to unoccupied Eu $4f$ state and (c) its magnified view indicated by dotted circle in (b), and (d) atomic orbital-resolved partial DOS.}
\label{fig4}
\end{figure}
Next, the electronic structure of the Eu-doped defect pyrochlore oxide \ce{KNbWO6}:0.125\ce{Eu^{3+}} (Dop model) is shown in Fig.~\ref{fig4}.
Due to different characteristics, we show the energy band structures and DOS for both the spin-up and spin-down states.
As shown in Fig.~\ref{fig4}(a), band gap was estimated to be the same to the Orig model as 3.00 eV, and an unoccupied band was found above the valence band maximum (VBM), denoted by dotted line.
To clarify which atom contributes to this unoccupied band, the square of band-decomposed wave function at the $\Gamma$ point of Brillouin zone (BZ) was plotted in Fig.~\ref{fig4}(b).
The isosurface of the wave function square was found mostly around Eu atom and partially around nearby O atoms (Fig.~\ref{fig4}(c)), indicating that this band is attributed to Eu atom.
In the orbital-resolved partial DOS (PDOS) plotted in Fig.~\ref{fig4}(d), some of spin-up Eu $4f$ states can be found at the position of $\sim3$ eV below VBM, corresponding to the six occupied $4f$ orbitals, and one spin-up Eu $4f$ state is seen above VBM, just being responsible for the unoccupied $4f$ orbital of \ce{Eu^{3+}} ion.
In the region of conduction band, spin-down $4f$ states and $5d$ states of Eu were found above CBM.
It should be noted that except Eu states, the other elements of host material have almost the same characteristics to the Orig model: VB mostly comes from O $2p$ states and CB is composed of the $4d/5d$ states of Nb/W atoms.

\begin{figure}[!t]
\centering
\includegraphics[clip=true,scale=0.18]{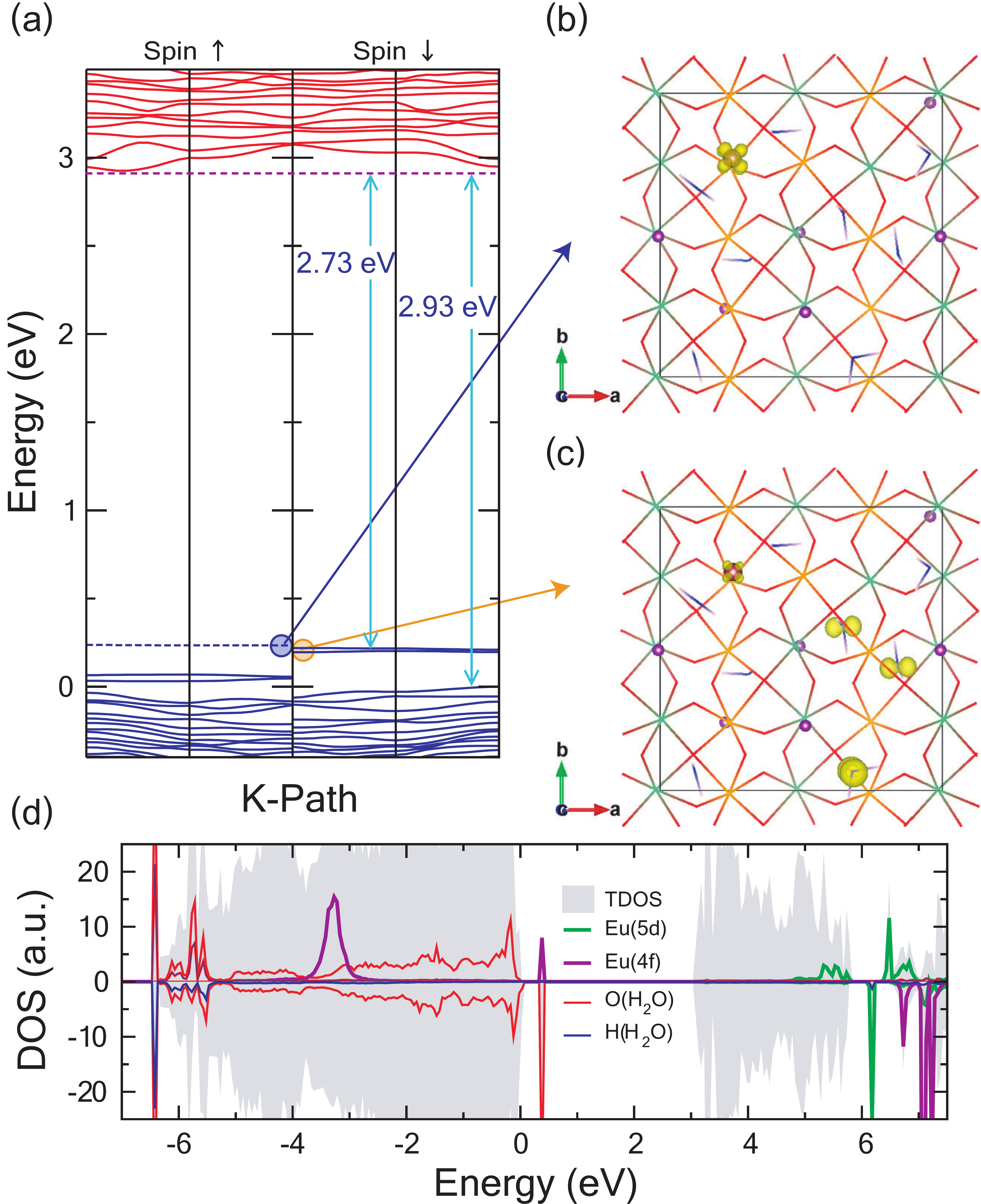}		
\caption{Electronic structure of hydrated Eu-doped defect pyrochlore oxide \ce{KNbWO6.H2O}:0.125\ce{Eu^{3+}} (HydDop model), calculated with PBEsol + $U$ method. (a) Energy band structure for spin-up and spin-down sates, isosurface view of wave function square at $\Gamma$ point for (b) unoccupied Eu $4f$ spin-up state and (c) occupied O $2p$ spin-down state of water molecule, and (d) atomic orbital-resolved partial DOS.}
\label{fig5}
\end{figure}
Figure~\ref{fig5} displays the electronic structure of the hydrated Eu-doped pyrochlore oxide \ce{KNbWO6.H2O}:0.125\ce{Eu^{3+}} (HydDop model), calculated with the PBEsol + $U$ method.
As shown in Fig.~\ref{fig5}(a) for its energy band structure, the band gap was calculated to be 2.73 eV, being 0.30 eV lower than the value (3.03 eV) of the undoped Hyd model, due to an up-shift of VBM attributed to two spin-down bands.
Like in the Dop model, one unoccupied spin-up band was found over VBM, corresponding to the unoccupied Eu $4f$ state, as shown in Fig.~\ref{fig5}(b) for the isosurface of band-decomposed wave function square at the $\Gamma$ point.
Meanwhile, the two occupied spin-down bands were revealed to be originated form oxygen atoms of water molecules, as demonstrated in Fig.~\ref{fig5}(c).
It is worth noting that such uplift of spin-down bands is not observed in the Hyd model, indicating that this is caused by interaction between the doped \ce{Eu^{3+}} ions and the water molecules.
Such interaction effect is more noticeable in atomic orbital-resolved PDOS analysis shown in Fig.~\ref{fig5}(d).
When compared with the dehydrated Eu-doped Dop model (Fig.~\ref{fig3}(a)), the spin-up Eu $4f$ states are found to be also divided into the occupied and unoccupied parts, but the unoccupied state is observed to be clearly up-shifted from the valence band of host material.
In addition, the location of spin-down Eu $4f$ states relative to spin-up ones is more or less the same to Dop model, but all the Eu $5d$ states move higher on energy scale.
When compared with the hydrated Hyd model (Fig.~\ref{fig3}(b)), the occupied O $2p$ states of water molecule are clearly shown to be up-lifted over the valence band of host material. 
To sum up, it can be concluded that the water molecules in the hydrated host material interact with doped \ce{Eu^{3+}} ion as well as the host compound, driving the non-radiative energy transfer between activator \ce{Eu^{3+}} ion and the host.

\subsection{Effect of oxygen vacancy on luminescence}
In general, points defects play a critical role in luminescence since they can act as traps for photo-generated electrons, leading to the non-radiative recombination~\cite{gonze6, walle}.
Considering that the oxygen vacancy $V_{\text{O}}$ is the dominant point defect with the lowest formation energy in oxides, we investigated only the effect of oxygen vacancy on luminescence.
Simulating oxygen vacancies with different charge states $V_{\text{O}}^q$ is not problem for the Orig and Hyd models: removing one oxygen atom and giving the charge $q$ to the O-removed model.
For the cases of Eu-doped models (i.e., Dop and HydDop), however, the conferred charge on the model could be captured by \ce{Eu^{3+}} ion (becoming \ce{Eu^{2+}} ion) rather than by oxygen vacancy $V_{\text{O}}$.
To verify the charge state of \ce{Eu} ion in existence of $V_{\text{O}}$, we calculated the band structures of \ce{KNbWO6}:\ce{Eu^{2+}} model and Dop+$V_{\text{O}}$ models with different total charges of $-1$, $0$, $+1$ and $+2$ (see Fig. S4).
It turned out that the band structures of the Dop+$V_{\text{O}}$ models with total charges of $-1$, $0$ and $+1$ have a characteristic feature of \ce{KNbWO6}:\ce{Eu^{2+}} $-$ extremely localized flat bands from the seven Eu $4f$ spin-up states in the middle of band gap region $-$ whereas the Dop+$V_{\text{O}}^{+2}$ model resembles that of \ce{KNbWO6}:\ce{Eu^{3+}}.
Therefore, the Dop+$V_{\text{O}}$ models with total charges of $-1, 0, +1$ and $+2$ represent the defect states of \ce{Eu^{2+}}/$V_{\text{O}}^0$, \ce{Eu^{2+}}/$V_{\text{O}}^{+1}$, \ce{Eu^{2+}}/$V_{\text{O}}^{+2}$ and \ce{Eu^{3+}}/$V_{\text{O}}^{+2}$, respectively.

We estimated the formation energies of these oxygen vacancies using the following equation~\cite{walle},
\begin{equation}
E_f(V_{\text{O}}^q) = E(\text{perf}+V_{\text{O}}^q) - E(\text{perf}) + \mu_{\text{O}} + q\varepsilon_{\text{F}} + E_{\text{MP}}
\label{eq1}
\end{equation}
where $E(\text{perf}+V_{\text{O}}^q)$ and $E(\text{perf})$ are the total energies of compounds with and without $V_{\text{O}}^q$, $\mu_{\text{O}}$ the chemical potential of oxygen, $\varepsilon_{\text{F}}$ the Fermi energy and $E_{\text{MP}}$ the Makov-Payne correction term for the finite size effect of charged model.
Here, $\mu_{\text{O}}$ was estimated to be half the total energy of isolated oxygen molecule, and $\varepsilon_{\text{F}}$ could be defined referencing to the VBM of the host as $\varepsilon_{\text{F}}=\varepsilon_{\text{VBM}}+\Delta\varepsilon_{\text{F}}$, where $\Delta\varepsilon_{\text{F}}$ is varying between 0 and band gap $E_{\text{g}}$~\cite{Kye18jpcl, Kye19jmcc}.
Table~\ref{tab3} lists the calculated formation energies $E_f(V_{\text{O}}^q)$ by using VBM as $\varepsilon_{\text{F}}$.
It was revealed that in each case of different models, $V_{\text{O}}^{+2}$ had the lowest formation energy, indicating that it is the oxygen vacancy with the highest possibility in the \ce{KNbWO6}-derived compounds, regardless of being hydrated or Eu doped.
The Eu$^{2+}$ doping was found to make it easy for oxygen vacancy to be generated due to lowering the formation energy when Eu$^{2+}$ doping in most cases, by comparing between Orig and Dop or Hyd and HydDop.
It should be highlighted that when compared with \ce{Eu^{2+}}/$V_{\text{O}}^{+2}$, \ce{Eu^{3+}}/$V_{\text{O}}^{+2}$ has remarkably higher formation energy, indicating that with oxygen vacancy the charge state of \ce{Eu^{3+}} is readily reduced to \ce{Eu^{2+}} with a half-filled electron configuration.
Above discussion shows that O vacancies in this compound may act as a donor to reduce \ce{Eu^{3+}} to \ce{Eu^{2+}}. 
In fact, similar mechanism by Ba vacancy defect (\ce{V_{Ba}}), acting as a donor, was proposed to explain the reduction of \ce{Eu^{3+}} to \ce{Eu^{2+}} observed in \ce{BaAl2O4$:$Eu} phosphor prepared under thermal carbon-reducing condition.~\cite{peng}
Once the \ce{Eu^{3+}} is reduced to \ce{Eu^{2+}}, it would exhibit characteristic broad emission spectra at around 500 nm due to $4f-5d$ transition, in contrast to sharp peaks in emission spectra of \ce{Eu^{3+}}.
However, in our previous experiment, the absorption and emission spectra were found to be originated from the intra-configurational $4f-4f$ transitions of \ce{Eu^{3+}}, and no characteristics of $4f-5d$ transitions of \ce{Eu^{2+}} was found~\cite{euHan}.
Thus, once the doped \ce{Eu^{3+}} ion is reduced to \ce{Eu^{2+}}, it can no longer act as luminescence centre in this material, indicating that the oxygen vacancies are detrimental to luminescence.
It is worth noting that all the $E_f(V_{\text{O}}^q)$ values in the hydrated compounds are higher than in case of dehydrated ones, indicating that water molecules inside the host structural framework suppress the formation of oxygen vacancies.
\begin{table}[!t]
\small
\caption{Formation energies of oxygen vacancies with different charge states ($V_{\text{O}}^q$) and transition energies between the charge states.}
\label{tab3}
\begin{tabular}{lcclcc}
\hline
& Orig & Hyd & & Dop & HydDop \\ 
\hline
\multicolumn{6}{l}{\quad Formation energy (eV)} \\
$V_{\text{O}}^0$ & 5.30 & 5.62 & \ce{Eu^{2+}}/$V_{\text{O}}^0$ & 5.36 & 5.54  \\
$V_{\text{O}}^{+1}$    & 3.02 & 3.32 & \ce{Eu^{2+}}/$V_{\text{O}}^{+1}$ & 2.50 & 2.67  \\
$V_{\text{O}}^{+2}$ & 1.28 & 1.76 & \ce{Eu^{2+}}/$V_{\text{O}}^{+2}$ & 0.01 & 0.28  \\
  &      &      & \ce{Eu^{3+}}/$V_{\text{O}}^{+2}$ & 1.35 & 1.53  \\
\hline
\multicolumn{6}{l}{\quad Charge state transition energy (eV)} \\
$\varepsilon(2+/1+)$ & 1.74 & 1.56 &  $\varepsilon(2+/1+)$(\ce{Eu^{2+}}) & 2.49 & 2.39  \\
$\varepsilon(2+/0)$  & 2.00 & 1.93 &  $\varepsilon(2+/0)$(\ce{Eu^{2+}})  & 2.68 & 2.63  \\
$\varepsilon(1+/0)$  & 2.27 & 2.29 &  $\varepsilon(1+/0)$(\ce{Eu^{2+}})  & 2.86 & 2.87  \\
\hline
\end{tabular}
\end{table}

\begin{figure}[!t]
\centering
\includegraphics[clip=true,scale=0.5]{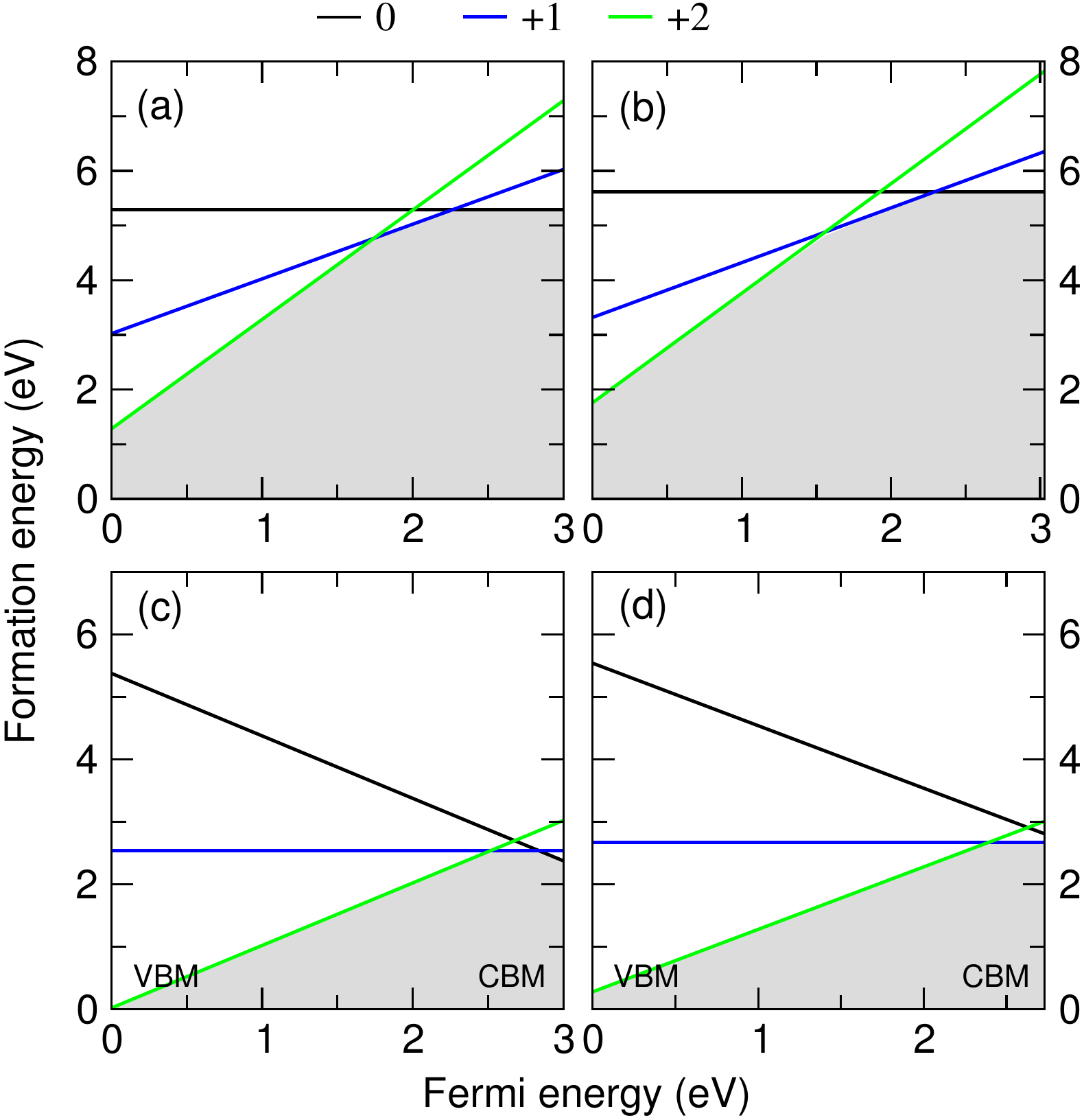}
\caption{Formation energy of oxygen vacancy with various charge states of $-$1, 0, +1 and +2 in (a) defect pyrochlore oxide \ce{KNbWO6} (Orig), (b) hydrated one \ce{KNbWO6.H2O} (Hyd), (c) Eu-doped one \ce{KNbWO6}:0.125Eu$^{3+}$ (Dop) and (d) hydrated Eu-doped one \ce{KNbWO6.H2O}:0.125Eu$^{3+}$ (HydDop), as functions of Fermi energy.}
\label{fig6}
\end{figure}

For the charged oxygen vacancies, their formation energies vary with change in the chemical potential of electron reservoir $\varepsilon_F$ in Eq.~\ref{eq1}.
Figure~\ref{fig6} presents the formation energies of oxygen vacancies in the models as a function of Fermi energy.
The charge state of oxygen vacancy with the lowest formation energy was found to change as the Fermi energy increases.
When $E_f(V_{\text{O}}^q)=E_f(V_{\text{O}}^{q'})$ at certain energy level, the charge state of $V_{\text{O}}$ may change from $V_{\text{O}}^q$ to $V_{\text{O}}^{q'}$ or vice versa by trapping or releasing the activated electron, and this energy level is defined as charge state transition energy $\varepsilon(q/q')$.
In the bottom part of Table~\ref{tab3}, the transition energies estimated by inspecting Fig.~\ref{fig6} were listed.
It was clearly revealed that the transition energies in the Eu-doped models were found in the higher Fermi energy close to the CBM, whereas they were located in the middle of band gap region.
Importantly, $\varepsilon$(2+/0) were estimated to be 2.68 and 2.63 eV in the Dop and Hyddop models, which are close to the transition energy 2.67 eV of \ce{^7F_0}$\rightarrow$\ce{^5D_2} being responsible for major peak of 464 nm in absorption spectra~\cite{euHan}.
This indicates that the oxygen vacancy in the Dop and HydDop models can act as a trap for electrons excited by incident radiation of 464 nm.
%

\subsection{Ln$^{3+}$ doping into \ce{KNbWO6} (Ln = Ce, Pr, Nd, Pm, Sm)}
Finally, we investigated the crystalline and electronic structures of other trivalent Ln ion doped pyrochlore oxides \ce{KNbWO6}:0.125Ln$^{3+}$ (Ln = Ce, Pr, Nd, Pm, Sm) with a calculation of doping energy.
As mentioned above, when the \ce{Ln^{3+}} ion is doped into the pyrochlore oxide host \ce{KNbWO6}, three \ce{K+} cations should be removed from the host to satisfy the charge neutrality.
Then, the doping process can be thought to occur through (1) taking away three K atoms from the Orig model, leading to formation of intermediate phase, and (2) inserting one Ln atom into the intermediate phase to reach the Ln-doped compounds.
For instance, the binding energies between the intermediate phase and three K atoms and one Eu atom can be calculated as follows, 
\begin{eqnarray}
E_b(\text{3K}) = E(\text{int}) + 3E(\ce{K}) - E(\text{Orig}) \\
E_b(\ce{Eu}) = E(\text{int}) + E(\ce{Eu}) - E(\text{Dop})
\label{eq2}
\end{eqnarray}
where $E(\text{int})$, $E(\text{Orig})$ and $E(\text{Dop})$ are the total energies of the crystalline compounds in the intermediate phase, Orig and Dop model, and $E(\ce{K})$ and $E(\ce{Eu})$ are the total energies of isolated K and Eu atoms, respectively.
Then the doping energy can be calculated as follows,
\begin{equation}
\begin{split}
E_{\text{dop}} & = E(\text{Dop}) - E(\ce{Eu}) + 3E(\ce{K}) - E(\text{Orig}) \\ 
& = E_b(\ce{3K}) - E_b(\ce{Eu})
\end{split}
\label{eq3}
\end{equation}
Likewise, the doping energy in case of the hydrated compound can be estimated from total energies of Hyd model $E(\text{Hyd})$ and HydDop model $E(\text{HydDop})$.

\begin{table}[!t]
\small
\caption{Lattice constants ($a,b,c$), unit cell volume and doping energy of \ce{KNbWO6}:0.125\ce{Ln^{3+}} (Ln = Ce, Pr, Nd, Pm, Sm, Eu), calculated by using PBEsol + $U$ method.}
\label{tab4}
\begin{tabular}{cccccc}
\hline
\multirow{2}{*}{Dopant} & \multicolumn{3}{c}{Lattice constants (\AA)} & Volume & $E_{\text{dop}}$ \\
\cline{2-4}
 & $a$ & $b$ & $c$ & (\AA$^3$) & (eV) \\
\hline
Ce & 10.39 & 10.34 & 10.38 & 1115.82 & 0.10 \\
Pr & 10.39 & 10.34 & 10.38 & 1114.60 & 0.13 \\
Nd & 10.39 & 10.33 & 10.37 & 1113.72 & 0.19 \\
Pm & 10.39 & 10.33 & 10.37 & 1112.83 & 0.28 \\
Sm & 10.38 & 10.33 & 10.37 & 1112.04 & 0.39 \\
Eu & 10.38 & 10.33 & 10.36 & 1111.31 & 0.48 \\
\hline
\end{tabular}
\end{table}
\begin{figure}[!th]
\centering
\includegraphics[clip=true,scale=0.5]{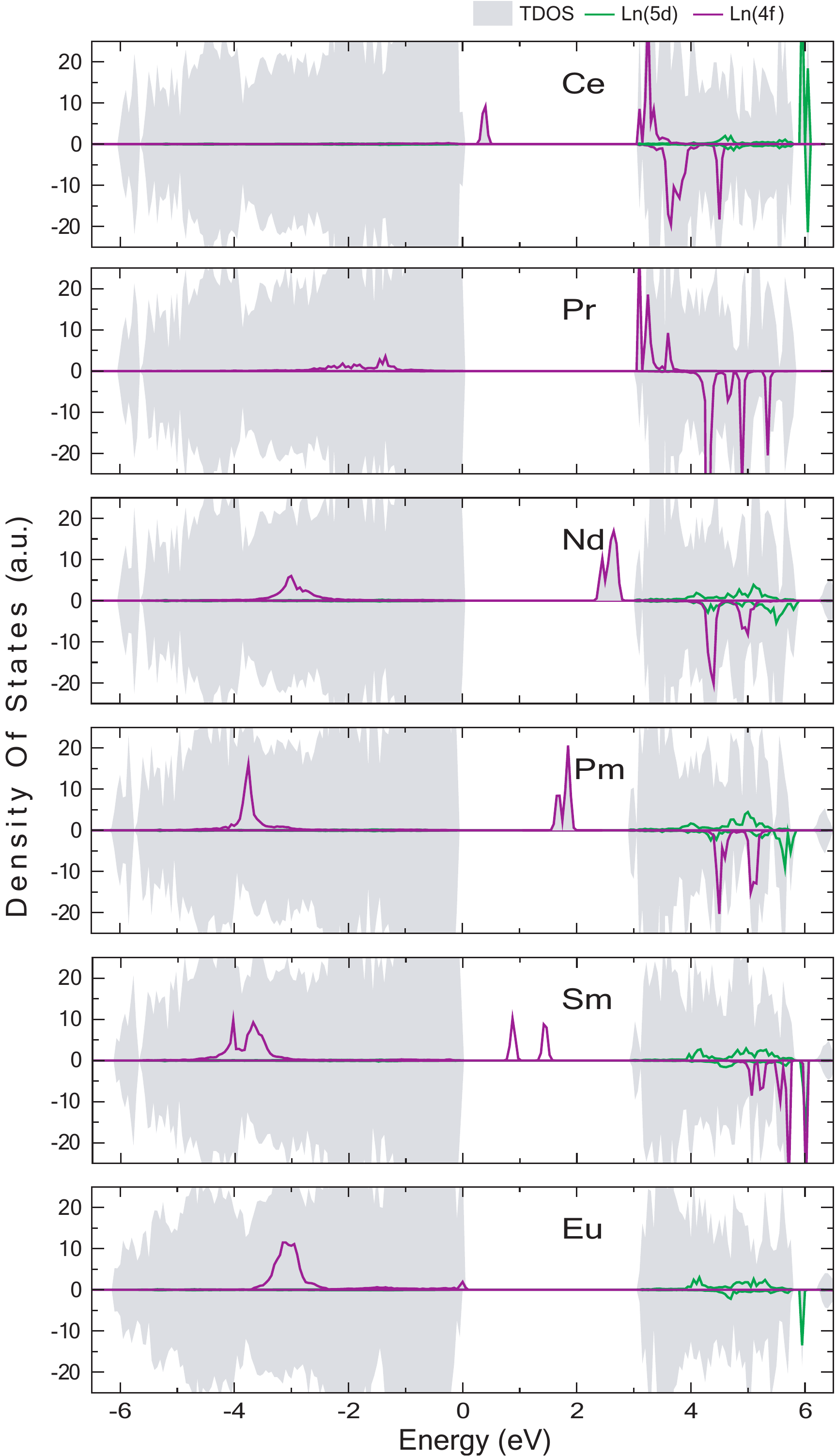}		
\caption{Partial density of states in Ln-doped pyrochlore oxides \ce{KNbWO6}:0.125\ce{Ln^{3+}} (Ln = Ce, Pr, Nd, Pm, Sm, Eu), calculated by using PBEsol + $U$ method.}
\label{fig7}
\end{figure}
Table~\ref{tab4} presents the optimized lattice constants and unit cell volumes of the Ln-doped pyrochlore oxides \ce{KNbWO6}:0.125Ln$^{3+}$ (Ln = Ce, Pr, Nd, Pm, Sm, Eu) and their doping energies.
As increasing the atomic number of Ln dopant ion, the lattice constants and thus the unit cell volume were found to gradually decrease; for the unit cell volume from 1115.82 \AA$^3$ for Ce- to 1111.04 \AA$^3$ for Eu-doped compounds.
Such gradual decrease in the unit cell volume can be attributed to decrease of the ionic radius of \ce{Ln^{3+}} ion as increasing the atomic number.
The doping energy was also found to systematically change; it increases from 0.10 eV for Ce-doping to 0.48 eV for Eu-doping.
This indicates that doping Ln into \ce{KNbWO6} is getting more difficult going from Ce to Eu.
Conversely, \ce{KNbWO6} can be said to serve as a better host material for \ce{Ce^{3+}} doping than for \ce{Eu^{3+}}.
For the case of Eu-doping into the Hyd model, the doping energy was calculated to be 0.57 eV, which is higher than into the Orig model (0.48 eV), indicating that the doping is more favorable in the dehydrated compound. 

Figure~\ref{fig7} shows the atomic orbital-resolved PDOS in the Ln-doped pyrochlore oxides, highlighting the PDOS of Ln $5d$ and $4f$ states in the grey background of total DOS.
In all the Ln-doped compounds, the band gap identified by TDOS was found to barely change for different Ln ions but to be fixed at $\sim$3 eV.
Commonly, the \ce{Ln^{3+}} ions contain seven $4f$ orbitals, which are divided into occupied and unoccupied states.
The number of occupied states corresponding to the spin-up electrons increases from 1 for \ce{Ce^{3+}} to 6 for \ce{Eu^{3+}}, while the unoccupied orbitals are for mostly spin-down states and some spin-up states.
In the case of \ce{Ce^{3+}}-doped compound, the occupied spin-up state was found over the VBM, but in other cases they were located below the VBM.
On the other hand, the unoccupied states were found above the CBM in the cases of \ce{Ce^{3+}}- and \ce{Pr^{3+}}-doped compounds, whereas they were found to gradually move towards the VBM going from \ce{Nd^{3+}} through \ce{Pm^{3+}} to \ce{Sm^{3+}}.
Even, the unoccupied spin-up $4f$ state of Eu is located just above the VBM, as discussed above.
It should be noted that as increasing the atomic number of Ln ion, all the spin-up $4f$ states move downward, while the unoccupied spin-down $4f$ states move upward.
Such tendency of $4f$ state positioning was observed by first-principles study on the series of \ce{Ln^{3+}}-doped \ce{LaSi3N5} (Ln = Ce, Pr, Nd, Pm, Sm, Eu) luminescent materials~\cite{ismail}.
These imply that the defect pyrochlore oxides \ce{KNbWO6} doped with \ce{Ln^{3+}} can also exhibit luminescent property, and therefore, further experimental studies are required to verify their potentialities as luminescent materials.

\section{\label{sec-conc}Conclusions}
In this study, we have carried out the first-principles calculations to clarify the several factors affecting luminescence properties of defect pyrochlore oxide \ce{KNbWO6}-based materials.
Several unit cell models have been suggested for crystalline \ce{KNbWO6}, its hydrated phase \ce{KNbWO6.H2O}, Eu-doped phase \ce{KNbWO6}:0.125Eu$^{3+}$ and hydrated Eu-doped phase \ce{KNbWO6.H2O}:0.125Eu$^{3+}$, in consideration of random distribution of Nb/W atoms, water molecule and Eu positions. 
Structural optimizations revealed that the local symmetry of \ce{Eu^{3+}} ion can be lowered in the dehydrated phase, giving the explanation for the experimental result of 10 times weaker emission at 612 nm owing to electric-dipole \ce{^5D0$-$^7F2} by annealing.
Through the electronic structure calculations, the band gap of the host has been found to be almost the same as $\sim$3 eV in all the \ce{KNbWO6}-derived compounds.
From the calculated DOS of \ce{KNbWO6$:0.125$Eu^{3+}}, the occupied spin-up $4f$ states of \ce{Eu^{3+}} ion were found below the VBM of the host, while the unoccupied spin-up state was found above the VBM.
The observed up-shift of the unoccupied spin-up state from the VBM in \ce{KNbWO6.H2O$:0.125$Eu^{3+}} indicates that water molecules could interact with host and \ce{Eu^{3+}} ion, mediating the non-radiative energy transfer between them and causing the quenching effect.
When introducing the oxygen vacancy, \ce{Eu^{3+}} ion was found to be readily reduced to \ce{Eu^{2+}}, leading to detriment to luminescence property, and moreover $V_{\text{O}}$ was found to act as traps for the electrons activated by incident light of 464 nm.
Finally the crystalline and electronic structures of \ce{KNbWO6} doped with other Ln ions (Ln = Ce, Pr, Nd, Pm, Sm) have been investigated, concluding that \ce{KNbWO6} can also be used as luminescent host for doping a series of \ce{Ln^{3+}} ions. 

\section*{Acknowledgments}
This work is supported as part of the basic research project ``Design of Innovative Functional Materials for Energy and Environmental Application'' (No. 2016-20) by the State Committee of Science and Technology, DPR Korea.
Computation was done on the HP Blade System C7000 (HP BL460c) that is owned by Faculty of Materials Science, Kim Il Sung University.

\bibliographystyle{elsarticle-num-names}
\bibliography{Reference}

\end{document}